\documentclass{aa}  

\usepackage{graphicx}
\usepackage{amsmath,amsfonts,amssymb}
\usepackage{aas_macros}
\usepackage[usenames]{xcolor}
\usepackage{caption,subfloat}
\usepackage[english]{babel}
\usepackage{gensymb}

\usepackage{txfonts}
\usepackage[colorlinks,allcolors=blue]{hyperref}

\begin{document}

   \title{Dusty spirals triggered by shadows in transition discs}

   \author{N. Cuello,
          \inst{1}\fnmsep\inst{2}
          M. Montesinos,
          \inst{3}\fnmsep\inst{4,8}
          S.M. Stammler,
          \inst{1}\fnmsep\inst{5}
          F. Louvet
          \inst{6}
          \and
          J. Cuadra\inst{1,2,4,7}
          }

   \institute{Instituto de Astrof\'isica, Pontificia Universidad Cat\'olica de Chile, Santiago, Chile\\
   		\email{ncuello@astro.puc.cl}
   \and
	Millennium Nucleus ``Protoplanetary discs'', Chile
   \and
        Instituto de F\'isica y Astronom\'ia, Universidad de Valpara\'iso, Av. Gran Breta\~na 1111, 5030 Casilla, Valpara\'iso, Chile 
	\and
	N\'ucleo Milenio de Formaci\'on Planetaria (NPF), Chile
	\and
	University Observatory, Ludwig Maximilian University, Scheinerstra{\ss}e 1, 81679 Munich, Germany	   
        \and
        Departamento de Astronom\'ia de Chile, Universidad de Chile, Santiago, Chile     
        \and 
        Max-Planck-Institut f\"ur extraterrestriche Physik (MPE), D-85748 Garching, Germany
        \and
        Chinese Academy of Sciences South America Center for Astronomy, National Astronomical Observatories, CAS, Beijing, China
     }

  \authorrunning{N. Cuello et al.}

   \date{Received \dots; accepted \dots}

% \abstract{}{}{}{}{} 
% 5 {} token are mandatory
 
  \abstract
   {Despite the recent discovery of spiral-shaped features in protoplanetary discs in the near-infrared and millimetric wavelengths, there is still an active discussion to understand how they formed. In fact, the spiral waves observed in discs around young stars can be due to different physical mechanisms: planet/companion torques, gravitational perturbations or illumination effects.}
   {We study the spirals formed in the gaseous phase due to two diametrically opposed shadows cast at fixed disc locations. The shadows are created by an inclined non-precessing disc inside the cavity, which is assumed to be optically thick. In particular, we analyse the effect of these spirals on the dynamics of the dust particles and discuss their detectability in transition discs.}
   {We perform gaseous hydrodynamical simulations with shadows, then we compute the dust evolution on top of the gaseous distribution, and finally we produce synthetic ALMA observations of the dust emission based on radiative transfer calculations.}
   {Our main finding is that mm- to cm-sized dust particles are efficiently trapped inside the shadow-triggered spirals. We also observe that particles of various sizes starting at different stellocentric distances are well mixed inside these pressure maxima. This dynamical effect would favour grain growth and affect the resulting composition of planetesimals in the disc. In addition, our radiative transfer calculations show spiral patterns in the disc at 1.6 $\mu$m and 1.3 mm. Due to their faint thermal emission (compared to the bright inner regions of the disc) the spirals cannot be detected with ALMA. Our synthetic observations prove however that shadows are observable as dips in the thermal emission.}
   {}


   \maketitle
%
%-------------------------------------------------------------------

\section{Introduction}
\label{sec:intro}

The process of star formation, through the collapse of a prestellar core, often leads to the creation of a protoplanetary disc made of gas and dust orbiting the proto-star \citep{Williams-Cieza-2011}. In the current paradigm of planet formation, protoplanetary discs are the birthplace of the first rocky bodies, also called planetesimals. These solids eventually become planets, either by core-accretion or by gravitational instabilities, in time-scales shorter than several Myrs \citep{Ribas+2014}. Up-to-date, more than $3\,500$ confirmed exoplanets have been detected\footnote{www.exoplanet.eu}, which clearly shows that planets are a common by-product of star formation. However, despite the large amount of theoretical and experimental studies in the field, the process by which planets form from micron-sized dust grains remains elusive \citep[and references therein]{Birnstiel+2016}.
	
In fact, there are several barriers for planet formation such as the radial-drift \citep{W77, Laibe+2012}, the bouncing \citep{Zsom+2010} and the fragmentation \citep{Blum&Wurm2008} of solids during their evolution inside the disc. These physical processes severely compromise the survival and/or the growth of such solids, and therefore prevent planet formation. Several mechanisms have been proposed to circumvent this problem: the presence of planetary traps \citep{Paardekooper&Mellema2004, Fouchet+2007, Pinilla+2012}, streaming instabilities \citep{Youdin&Goodman2005, Jacquet+2011, Johansen+2012}, anticyclonic vortices \citep{Barge&Sommeria1995, Meheut+2012, Baruteau&Zhu2016}, dead zones \citep{Kretke&Lin2007, Dzyurkevich+2010}, meridional circulation \citep{Fromang+2011}, photophoretic transport \citep{Krauss&Wurm2005, Cuello+2016}, radiation pressure effects \citep{Vinkovic2014} and self-induced dust traps \citep{Gonzalez+2017} among others. The common denominator of these processes is the concentration and/or the growth of dust at a particular location of the disc, either by stopping the radial-drift or by lowering the relative velocities of collision among solids. The former avoids solids to be accreted by the central star, while the latter allows particles to grow and eventually decouple from the gaseous phase.
	
In this work, we study the effect of shadow-triggered gaseous spirals on the dust distribution of a transition disc. In general, spiral wakes can be caused by different mechanisms: planet torques \citep{GoldreichTremaine1979}, gravitational instabilities \citep[and references therein]{KratterLodato2016}, stellar encounters \citep{Pfalzner2003} and illumination effects \citep{Montesinos+2016}. Regardless of their origin, these perturbations cause a local symmetry breaking in the disc, which has profound consequences for structure evolution and planetesimal formation. In the recent years, due to the spectacular increase in angular resolution and the development of extreme adaptive optics, it has been possible to reveal astonishing asymmetric features in several transition discs. These observations can be split in two kinds according to the wavelength: the near-infrared (NIR) or scattered-light ones, which trace the small micron-sized particles; and the (sub-)millimetric ones, which map the thermal emission of large mm/cm dust grains. In Table~\ref{tab:obs}, we report some of the most interesting transition discs exhibiting visible spirals in the NIR and/or large asymmetric dust traps (e.g. HD~142527 and HD~135344B).
	
Elias~2-27 is a system of particular interest. In fact, \cite{Perez+2016} report a couple of symmetric spirals in the dust thermal emission at 1.3 mm with ALMA. To our knowledge, this is the only detection of a grand-design spiral in the large grains distribution, i.e. close to the mid-plane. It remains however unclear whether the spirals and the shadows detected by \cite{Stolker+2016} and by \cite{Benisty+2017} are related to planet formation processes or not. For instance, in the protoplanetary disc around the F-type star HD~135344B, \cite{Stolker+2016} report the detection of four shadows (A, B, C and D) in the J-band and at least three spirals arms. Surprisingly, shadow C is not detected in earlier observations (R-, I- and Y-bands), which seems to indicate its transient nature. This could be explained by a local perturbation in the inner regions of the disc or by an accretion funnel flow from the inner disc onto the star. In principle, an outer planet could excite a pair of symmetrical spirals as suggested by \cite{Dong+2015}, even though there is no evidence of planetary gaps whatsoever. Until now, the only works that established a clear connexion between shadows and spirals are \cite{Montesinos+2016} and \cite{Montesinos&Cuello2018}. However, it is still unknown if this mechanism can explain the shadows in HD~135344B and how it affects the dust distribution in the disc.
	
After the detection of dusty spirals in Elias 2-27 by \cite{Perez+2016}, many scenarios have been proposed and tested to explain the formation of such spirals. For instance, \cite{Tomida+2017} report similar spirals in fairly massive three-dimensional discs (class-I objects) formed via magnetohydrodynamic (MHD) effects during the phase of envelope collapse. In this case, the pair of grand-design spiral arms form due to gravitational instabilities in the disc. Interestingly, the spirals disappear in a few rotations and new spirals form due to growing instabilities in the gas. This spiral formation process occurs throughout the class-0 and class-1 phases. \cite{Meru+2017} also study gravitationally unstable discs and show that it is possible to create spirals analogue to those of Elias 2-27, without considering MHD effects. In addition, the authors also consider the case of an internal (<~$300$ au) and an external (>~$300$ au) companion. The former configuration fails to produce spirals as the ones observed, while the latter is able to create grand-design spirals in the disc. The companion must however be comprised between $300$ and $700$ au and with a mass below $13$ Jupiter masses. Both scenarios, gravitational instability and the presence of an exterior companion, seem equally likely with the current observational constraints.

In this paper, we consider the formation of spirals triggered by shadows proposed by \cite{Montesinos+2016}, which form in the presence of light-obstructing material close to the star \citep{Marino+2015}. This mechanism is due to the peculiar azimuthal acceleration caused by the shadows cast at the inner rim of the disc (cf. figure~3 in \cite{Montesinos+2016}). In this scenario, the spirals in the gaseous disc have lifetimes of the order of several thousand years and could in principle be detected in the NIR (\textit{H} band, 1.6 $\mu$m). However, we are left with the following questions: can these shadow-triggered spirals efficiently trap dust? If yes, what are the observational signatures of this mechanism? It is not trivial to address these questions because dust dynamics strongly depends on particle size, intrinsic density and gaseous local conditions (cf. Sec.~\ref{sec:dustdynamics}).

 \begin{table}[h]
 \centering
 \caption[]{\label{tab:obs}Detections of spiral and asymmetric dust traps in transition discs.}
\begin{tabular}{ccc}
\hline \hline
  System &
  Scattered-light &
  Continuum emission \\
 \hline
	HD~142527 & 1 & 2 \\
	MWC~758 & 3, 4  & - \\
	HD~100453 & 5, 6 & -  \\
	HD~135344B & 7, 8 & 9 \\
	Oph~IRS~48 & 10 & 11 \\ 
	HD~100546 & 12 & - \\
	Elias~2-27 & - & 13 \\	
\hline
\end{tabular}
\tablebib{(1)~\citet{Christiaens+2014}; (2)~\citet{Casassus+2015}; (3)~\citet{Grady+2013}; (4)~\citet{Benisty+2015}; (5)~\citet{Wagner+2015}; (6)~\citet{Benisty+2017}; (7)~\citet{Muto+2012}; (8)~\citet{Stolker+2016}; (9)~\citet{vanderMarel+2016}; (10)~\citet{Follette+2015}; (11)~\citet{vanderMarel+2013}; (12)~\citet{Garufi+2016}; (13)~\citet{Perez+2016}.
}
\end{table}

The radial motion of dust particles embedded inside a gaseous sub-Keplerian disc, i.e. in presence of radial pressure gradient, has been long known and extensively studied \citep{W77, NSH1986, Laibe+2012}. More recently, the case of an azimuthal pressure gradient and the resulting dust dynamics was considered by \cite{Birnstiel+2013}. In both cases, solid particles tend to move towards the pressure maxima at velocities which depend on their Stokes number and the drag regime (cf. Section~\ref{sec:eomdust}). So far, the dust behaviour in presence of both, radial and azimuthal, pressure gradients has been modestly explored, and only for a handful of mechanisms: planetary wakes \citep{PaardekooperMellema2006, Ayliffe+2012}, disc instabilities \citep{Rice+2004, Dipierro+2015, Tomida+2017, Meru+2017} and gravitational perturbations \citep{Zsom+2011, Dai+2015}. Thus, the dust dynamics in presence of the shadow-triggered gaseous spirals proposed by \cite{Montesinos+2016} remains to be studied.
	
The aim of this work is twofold: on the one hand, we analyse in detail the dust dynamics in presence of shadow-triggered spirals; and, on the other hand, we produce synthetic ALMA observations of the thermal emission of the dust. This allows us to address the two questions aforementioned about the dust trapping and its observability. The paper is structured as follows: in Section~\ref{sec:method}, we discuss the three main steps of our calculations; in Section~\ref{sec:results}, we report our results and the images obtained; in Section~\ref{sec:discussion}, we discuss our results; and, finally, we draw our conclusions in Section~\ref{sec:conclusions}.

%--------------------------------------------------------------------

\section{Numerical method}
\label{sec:method}

The approach followed in this work is based on three main steps, which connect our hydrodynamical simulations with the observations in a consistent way:
\begin{enumerate}
\item gaseous hydrodynamical simulations with shadows,
\item dust evolution on top of the gaseous distribution,
\item synthetic ALMA observations of the dust emission.
\end{enumerate}
The first hydrodynamical step, described in Section~\ref{sec:FARGO}, allows us to obtain the dynamical fields of the gas (velocity, temperature, density, acceleration) required for the dust evolution calculation. The equations of motion for the dust particles and their numerical integration are reported in Section~\ref{sec:dustcode}. Finally, based on the obtained dust distributions, we simulate the thermal emission of the dust and emulate ALMA observations of our system in Section~\ref{sec:images}.

	\subsection{Hydrodynamical simulations with shadows}
	\label{sec:FARGO}

We model the evolution of a self-gravitating circumstellar gaseous disc by means of the public two-dimensional hydrodynamic code {\sc fargo-adsg}\footnote{http://fargo.in2p3.fr/-FARGO-ADSG-} \citep{Masset-2000, Baruteau-Masset-2008}. The code {\sc fargo-adsg} solves the Navier-Stokes, continuity and energy equations on a staggered mesh in polar coordinates $(r,\theta)$. The simulations were performed with an improved version of this code, which includes physical mechanisms such as cooling, stellar heating and shadows. This allows us to simulate the evolution of a transition disc in presence of projected shadows as in \cite{Montesinos+2016}. These are considered as regions of the disc for which the incoming stellar irradiation is equal to 0. Below, we describe the disc model considered, the energy equation with shadows and the initialisation of the hydrodynamical simulations. The evolution of the disc is followed for about 10\,000 years.

    \subsubsection{Disc model}
    \label{sec:setup}

We consider a gaseous disc orbiting a solar-like star of mass $M_\star = 1 \, M_\odot$ and luminosity $L_\star = 1 \, L_\odot$. The disc extends from $r_{\rm in} = 100$ au to $r_{\rm out} = 600$ au. The total disc mass is set to 0.25 $M_\odot$ and its initial density profile is given by:
\begin{equation}
\label{eq:sigma}
	\Sigma(r) = 711 \, {\rm g \, cm^{-2}} \, \times \left( \frac{r}{\rm 1 \, au} \right) ^{-1} \,\,\, .
\end{equation}
This model corresponds to a marginally stable disc for which the Toomre parameter remains above $1$ throughout the simulation (cf. Appendix~\ref{sec:sanity-checks}).
The computational domain in physical units extends from $r_{\rm in}$ to $r_{\rm out}$ au over $n_r = 512$ logarithmically spaced radial cells. The grid samples $2 \pi$ in azimuth with $n_\theta = 1024$ equally spaced sectors. We also ran simulations at higher resolution with $n_r = 512$ and $n_\theta = 2048$ obtaining the same results, which means that the aforementioned resolution is accurate enough for our model. 
	
The turbulent viscosity of the gaseous disc is modelled by considering an $\alpha$-disc \citep{Shakura+Sunyaev1973}, with the $\alpha$ parameter set to $10^{-4}$. We assume a constant opacity equal to  $\kappa = 1 ~ \rm cm^2 g^{-1}$. It is worth noting that the opacity directly impacts on the cooling rate $Q_-$ of Eq.~\ref{eq:Qcool} in Sect.~\ref{sec:energyequation}. Interestingly, as shown in \citet{Montesinos+2016}, shadow-triggered spirals appear for high and low values of $\kappa$: 130 and 1 cm$^2$g$^{-1}$, respectively. A more realistic opacity depending on both the temperature and the density, e.g. the Rosseland mean opacity \citep{Semenov+2003}, would have a minor effect on the disc for the range of temperatures considered in this work. This justifies our choice of a constant opacity. The only heat source we consider is the stellar irradiation and we do not take into account the viscous heating, which is dominant only for the very inner regions of the disc (cf. Appendix~\ref{sec:heating}). All our models include self-gravity, meaning that the gas feels the gravitational potential of the star as well as of the disc itself. Given the mass of our system, this term cannot be neglected. Moreover, it has important effects for the dust evolution as discussed in Sec.~\ref{sec:eomdust}

\subsubsection{Energy equation and shadows model}
\label{sec:energyequation}

We implement the shadows in the disc in the same fashion as in \cite{Montesinos+2016}. If there are no shadows, the disc is irradiated by stellar light at all azimuths. The stellar irradiation heating per unit area is given by \cite{Frank-2002}:
\begin{equation}
\label{eq:Qs}
	Q^+_\star(r) = (1 - \beta) \frac{L_\star}{4 \pi r^2} \frac{H}{r} \left( \frac{{\rm d} \ln H}{{\rm d} \ln r} - 1  \right),
\end{equation}
where $\beta$ is the albedo, $L_\star$ the stellar luminosity and $H$ the disc scale height. We assume $\beta = 0$, which means that all the incoming radiation gets absorbed by the disc. Assuming local hydrostatic equilibrium, the scale height $H$ is given by:
\begin{equation}
	H =  \frac{c_{\rm s}}{v_{\rm k}} \, r \,\,\, ,
\end{equation}
where $c_{\rm s}$ is the sound speed and $v_{\rm k}$ the Keplerian velocity. During the calculation, the value of $H$ is continuously updated according to the evolution of the temperature of the disc. Also, the viscosity is given by $\nu =  \alpha  \, c_{\rm s} \, H$.

\begin{figure}
\centering
\includegraphics[width=8cm]{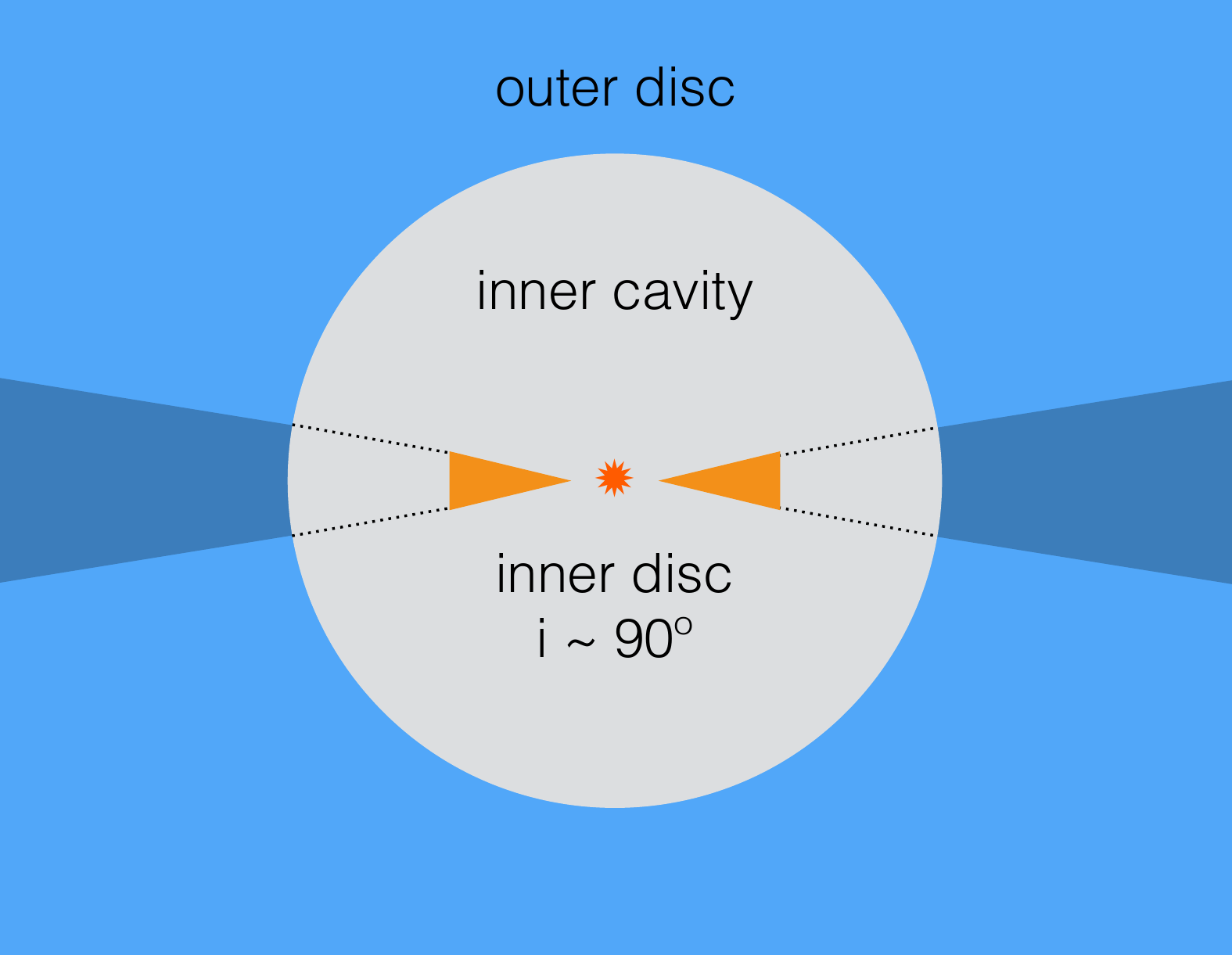}
\caption{Sketch of the inner regions of the disc model considered in this work. The outer disc is face-on, while the non-precessing inner disc is edge-on. The light-obstructing material around the star casts shadows at two diametrically opposed regions of the outer disc, in dark blue.}
\label{fig:disc}
\end{figure}	

We further assume the presence of a non-precessing edge-on disc ($i = 90\degree$) inside the large cavity. This type of structures can be created by an inner planetary or stellar companion as in \cite{Facchini+2018} and blocks the stellar light. In this configuration, two static and diametrically opposed shadows are cast onto the outer disc as shown in Figure~\ref{fig:disc}. In the case of prograde rotating shadows, with rotation periods smaller than 20 kyr, planetary-like spiral can appear at the corotation region with the shadows \citep{Montesinos&Cuello2018}. This effect is neglected in this work. In other words, our assumption of static shadows is valid for shadows with precession periods larger than roughly 20 kyr. It is worth highlighting that this inner disc is not treated in the hydrodynamical simulations, only its radiative effect are considered. Thus, we modify the stellar irradiation rate to account for the shadows:
\begin{equation}
\label{eq:QstarFinal}
	Q^+_d(r, \theta) = s(\theta) \, Q^+_\star (r) \,\,\, ,
\end{equation}
where the function $s(\theta)$ is given by:
\begin{equation}
\label{eq:sfunc}
	s(\theta) = 1 - A \exp{(-\theta^4/\sigma^2)} - A \exp{(-(\theta - \pi)^4/\sigma^2)} \,\,\, ,
\end{equation}
with $A=0.999$ and $\sigma=0.05$. This creates two shadows at $\theta = 0\degree$ and at $\theta = 90\degree$ with a width of about $29\degree$, where the stellar heating rate is set to nearly zero. The width is motivated by HD~142527's case as reported by \cite{Marino+2015}, and is also comparable with HD~135344B's shadow B, as reported by \cite{Stolker+2016}. By construction, the illuminated and the shadowed regions are smoothly connected.

Our hydro code solves the non-stationary equation for the thermal energy density, $e$, given by:
\begin{equation}
\label{eq:energy1}
	\frac{\partial e}{\partial t} + \mathbf{\nabla} \cdot (e \, \mathbf{v}) = -P \mathbf{\nabla} \cdot \mathbf{v} + Q^+_d  - Q^- \,\,\, ,
\end{equation}
where $\mathbf{v}$ is the gas velocity, $P$ the gas pressure, $Q^+_d$ the stellar heating rate per unit area (cf. Eq.~\ref{eq:QstarFinal}) and $Q^-$ the radiative cooling of the disc. For our calculations, we use:
\begin{equation}\label{eq:Qcool}
	Q^- = 2 \sigma_{\rm SB} T^4 / \tau\,\,\, ,
\end{equation}
where $\sigma_{\rm SB}$ is the Stefan-Boltzmann constant and $\tau$ the optical depth given by $\tau =\frac{1}{2} \kappa \Sigma$. In order to close the system of equations, we consider an ideal equation of state: $P = \Sigma \, T \, \mathcal{R}$, with $T$ the gas temperature and $\mathcal{R} = k_{\rm B}/\mu m_{\rm p}$, where $k_{\rm B}$ is the Boltzmann constant, $\mu$ the mean molecular weight of the gas and $m_{\rm p}$ the mass of the proton. Thus, the relationship between the thermal energy density and the temperature reads:
\begin{equation}
\label{eq:energyT}
	e = \Sigma \, T \frac{\mathcal{R}}{\gamma -1} \,\,\, ,
\end{equation}
where $\gamma$ is the adiabatic index. Finally, in order to avoid unrealistically low temperatures in the disc, we set the Cosmic Microwave Background (CMB) temperature ($2.7\,$K) as the floor temperature in our simulations.

		\subsubsection{Initialisation}
		\label{sec:initialization}

In a stationary regime, the energy received from the star equates the energy loss through the disc's surface. In order to reach a steady-state regime as fast as possible, the disc is initialised under the condition  $Q_\star^+ = Q^-$. Assuming vertical hydrostatic equilibrium, the gas temperature is related to the aspect-ratio of the disc as follows:
\begin{equation}
\label{eq:Temp}
	T(r) = \left(   \frac{H}{r}  \right)^2 \frac{G M_\star}{r} \frac{\mu}{\mathcal{R} } \,\,\, . 
\end{equation}

Defining $h = H/r = h_0 r^f$ and equating Eq. (\ref{eq:Qs}) with Eq. (\ref{eq:Qcool}), we obtain: 
\begin{equation}  
\label{eq:aspect}
	\frac{H}{r} = \left[  \frac{(1-\beta) L_\star}{16 \pi \sigma_{\rm SB}}    \frac{\Sigma_0 \kappa}{G^4 M_\star^4} \mathcal{R}^4  f \right]^{1/7} r^{f},
\end{equation}
where $f$ is the flaring index of the disc and is given by $f = \mathrm{d} \ln(H/r)/\mathrm{d} \ln r$. Eq. \ref{eq:aspect} gives a recipe to initialise the disc's aspect-ratio as a function of the initial disc and star properties. Formally, this equation is not self-consistent because the aspect-ratio depends on its own derivative. However, the dependence on the flaring index is very weak ($\propto f^{1/7}$). Therefore, as a first order approximation, we can assume that $f$ is nearly constant and equal to $f \simeq 1/7$.
 
%------------------------------------------------------------------------------------------------------------------------------

	\subsection{Dust evolution calculation}
	\label{sec:dustcode}

The outputs of our hydro simulation are used as inputs for our dust code, which computes the evolution of an ensemble of $N$ independent dust particles. Given that it is a post-processing calculation, the gaseous phase is not affected by the dust distribution. On the contrary, the dust is affected by the gas dynamics due to the aerodynamical drag, which strongly depends on the local disk conditions ($P$, $T$). It is useful to introduce the stopping time ($t_{\rm s}$) defined as the time it takes to the (Keplerian) dust to reach the same velocity as the (sub-Keplerian) gas. The dimensionless parameter called Stoked (noted St) is also relevant for dust dynamics since it measures the dust coupling to the gas, i.e. the drag regime. In the Epstein regime, valid for particles with sizes smaller than $9/4$ of their mean free path, St reads as follows:
\begin{equation}
\label{eq:St}
	{\rm{St}} = t_{\rm s} \, \Omega_{\rm K} = \frac{\rho_{\rm d} a}{\rho_{\rm g} c_{\rm s}} \, \Omega_{\rm K} \,\,\, ,
\end{equation}
where $\Omega_{\rm K}$ is the Keplerian frequency, $\rho_{\rm d}$ the bulk density of the particle, $a$ its radius, $\rho_{\rm g}$ the volumetric gas density and $c_{\rm s}$ the speed of the sound. There are three different regimes: for ${\rm St} \ll 1$, the dust is well-coupled to the gas; for ${\rm St} \sim 1$, the dust particles are marginally coupled to gas and feel the strongest radial-drift; and finally, for ${\rm St} \gg 1$, the particles are completely decoupled from the gas. Given the density profile considered (cf. Eq.~\ref{eq:sigma}), for a fixed grain size $a$, the Stokes number is an increasing function of the distance to the star.
	
	\subsubsection{Equations of motion and numerical integrator}
	\label{sec:eomdust}
    
Our dust code is based on previous work by \cite{Paardekooper-2007} and \cite{Zsom+2011}. In addition to the aerodynamical drag, dust particles feel both the gravitational potential from the star and from the gaseous disc. For a given dust particle of index $i$, we note ($r_{i}$, $\theta_{i}$) and ($v_{r, i}$, $v_{\theta, i}$) the polar coordinates and velocities respectively. Therefore, the equations of motion read as follows:
\begin{eqnarray}
	\frac{{\rm d}r_i}{{\rm d}t} &=& v_{r,i} \label{eq:eomdust1}\\
	\frac{{\rm d}\theta_i}{{\rm d}t} &=& \frac{L_i}{r_i^2} \label{eq:eomdust2} \\
	\frac{{\rm d}v_{r,i}}{{\rm d}t} &=& \frac{L_i^2}{r_i^3} - \frac{\partial \Phi}{\partial r} + F_{r,i} \label{eq:eomdust3}\\
	\frac{{\rm d}L_i}{{\rm d}t} &=& - \frac{\partial \Phi}{\partial \theta} + F_{\theta, i} \label{eq:eomdust4}
\end{eqnarray}
where $L_i$ is the specific angular momentum, and $F_{r,i}$ and $F_{\theta, i}$ are respectively the radial and azimuthal drag forces for particle $i$. $\Phi$ is the gravitational potential defined as: $\Phi = G M_* / R + \Phi_{\rm d}$, where $\Phi_{\rm d}$ is the disc's gravitational potential. The latter is given by the Fourier transform of the surface density $\Sigma$ of the disc. In order to avoid this computationally expensive calculation, we use the radial and azimuthal accelerations given by the hydro code. It is worth noting that this term increases the orbital velocity of the gas compared to non self-gravitating discs. Thus, it is important to include it because otherwise the particles in the outer parts of the disc would orbit too slowly compared to the gas. This would result in a constant and unrealistic tailwind, which would send dust particles outwards.

In general, Eqs.~\ref{eq:eomdust1}-\ref{eq:eomdust4} also have inertial forces that arise from the fact that the coordinate frame of the hydro code is always centred on the star, and not on the centre of mass. In the case of a binary system for instance, this introduces inertial forces that have to be taken into account. Given that our system is axisymmetric, the centre of mass is always at the centre of the star and we can therefore neglect these additional forces. The drag force on the dust particles is computed as follows:
\begin{equation}\label{eq:drag}
	\mathbf{F_{\rm drag}} = - \frac{\Omega_{\rm K}}{{\rm St}} \, \Delta {\mathbf{v}} \,\, ,
\end{equation}
where $\Delta {\mathbf{v}}$ is defined as the vectorial relative velocity between the dust, ${\mathbf{v}_{\rm dust}}$, and the gas, ${\mathbf{v}_{\rm gas}}$. The radial and azimuthal components of $\mathbf{F_{\rm drag}}$ are equal to $F_{r,i}$ and $F_{\theta, i}$ in Eqs.~\ref{eq:eomdust3} and \ref{eq:eomdust4} respectively. The expression of the Stokes number used in Eq.~\ref{eq:drag} is more general than the one in Eq.~\ref{eq:St}, which is only valid in the Epstein regime. In Appendix~\ref{sec:integration}, we report the method to compute the Stokes number in Eq.~\ref{eq:drag} and the details of the numerical integrator. We use a standard first-order leapfrog integrator to solve Eqs.~\ref{eq:eomdust1}-\ref{eq:eomdust4}, similar to that used by \cite{Paardekooper-2007}. In Appendix~\ref{sec:benchmark} we also detail the benchmark analysis of the dust evolution code, which ensures the accuracy of the numerical integrator.

It is worth highlighting that, in order to ensure an accurate integration of Eqs.~\ref{eq:eomdust1}-\ref{eq:eomdust4}, we perform interpolations between every two consecutive gas snapshots\footnote{as a rule of thumb, we require at least 10 hydro snapshots per orbit at the inner boundary.}. More specifically, we feed the dust code with the values of the required physical fields of the gas (i.e. $\Sigma$, $T$, $P$, ${\mathbf{v}_{\rm gas}}$ and ${\mathbf{a}_{\rm gas}}$). By doing so, we consistently take into account the evolution of the gas disc to compute the resulting motion of the dust particles.

	\subsubsection{Dust simulation setup}

For each dust simulation, we use 100\,000 dust particles with a fixed size between $a_{\rm min} = 1  \, \mu$m and $a_{\rm max} = \, 10$ cm. The particles are initially distributed according to the initial gas surface density of Eq.~\ref{eq:sigma}. The bulk density of the solids, $\rho_{\rm d}$, is set to 2.0~${\rm g\,cm^{-3}}$ to model a mixture of pyroxene, refractory organics and ices.
	
This distribution is a simplified initial condition since we are not considering a radial-dependent size distribution. The reason for this choice is that it allows us to better follow the dust kinematics by directly comparing the different grain species considered. For instance, in a more evolved protoplanetary disc, one would expect to have a more compact disc of millimetric grains compared to the one of micron-sized grains \citep{Laibe+2012}. The implications of this assumption are discussed in Sect.~\ref{sec:dustmixing}.

%-------------------------------------------------------------------------------------------------------------------------------

	\subsection{Radiative transfer and synthetic ALMA observations}
	\label{sec:images}

	\subsubsection{Radiative transfer with {\sc radmc-3d}}
	\label{sec:radiative-transfer}

Since our goal is to connect our simulations with ALMA observations, we need to compute the thermal emission of the dust distribution at millimetric wavelengths. To perform radiative transfer calculations we use the {\sc radmc-3d} sofware\footnote{http://radmc3d.ita.uni-heidelberg.de/} v0.41 \citep{radmc3d}. However, {\sc radmc-3d} input is a continuous three-dimensional dust density distribution, for each particle size in the radiative transfer model. Thus, we need a conversion from a discrete (cf. Section~\ref{sec:dustcode}) to a set of continuous dust distributions of different sizes. This is done in three main steps:

1) We spread all dust particles over a two-dimensional surface density grid with an exponential function:
\begin{equation}
	\Sigma_{{\rm dust},i}(\mathbf r) = \frac{m_i}{2 \, \pi D^2} \exp \left( \frac{\left( {\mathbf r} - {\mathbf r}_i  \right)^2}{2 \, D^2} \right) \,\, .
\end{equation}
$\Sigma_{{\rm dust},i}$ is the surface density that one particle of mass $m_i$ at location ${\mathbf r}_i$ contributes to the overall dust surface density. The factor $1/(\pi D^2)$ conserves the mass of the particle, but the overall surface density has to be gauged later nevertheless. As kernel size, we use $D=100$ au. This seems large compared to the size of our system, but as seen in Section~\ref{sec:spiralformation}, this is roughly the width of one spiral arm. Smaller values lead to noisy images.

2) We sort the particles into 12 particle size bins. Since we have particles between $a_{\rm min} = 1 \, \mu$m and $a_{\rm max} = 10 $ cm, this leads to 2 bins per decade. We then sum the individual single-particle surface densities in 12 total dust surface densities according their size:
\begin{equation}
	\Sigma_{{\rm dust},j} = \sum\limits_{\substack{i \\ a_{j,1} < a_i < a_{j,2} }} \Sigma_{{\rm dust},i} \,\, ,
\end{equation}
where $a_{j,1}$ and $a_{j,2}$ are the lower and the upper limits, respectively, of the size bin of index $j$.

3) Finally, these surface densities are vertically distributed with a Gaussian to get spatial mass densities according to:
\begin{equation}
	\rho_{{\rm dust}, j} ( {\mathbf r}, z ) = \frac{\Sigma_{{\rm dust}, j}({\mathbf r})}{\sqrt{2 \, \pi} h_j} \exp \left( - \frac{z^2}{2 h_j^2} \right) \,\, ,
\end{equation}
with the dust scale height given by \cite{Birnstiel+2010} as:
\begin{equation}
	h_j = H \cdot \min \left( 1, \, \sqrt{ \frac{\alpha}{ \min({\rm St}_j, 0.5) (1+{\rm St}_j^2) } } \right) \,\, .
\end{equation}
The combined mass of all our $100\,000$ dust particles is far too low for a typical disc. Thus, to obtain a dust-to-gas $\sim 1\%$, we multiply our densities with a constant factor such that the total mass of the dusty disc is $M_{\rm dust} = 0.25 \times 10^{-2} \, M_{\odot}$.
	
We feed {\sc radmc-3D} with the dust temperature and density distribution obtained from the hydro and the dust evolution simulations, respectively. The radiative transfer is then performed through the Monte Carlo method of {\sc radmc-3d} based on Mie theory \citep{Bohren+Huffman1983}. We use the optical constants taken from \cite{Dorschner+1995} assuming the particles are spherical and consist of pure pyroxene (${\rm Mg_{0.7}Fe_{0.3} SiO_3}$)). It is worth highlighting that we modified {\sc radmc-3d}, such that no photons are sent from the star within two sectors of angle $15\degree$ at azimuth $0\degree$ and $180\degree$ as shown in Fig.~\ref{fig:disc}. This mimics the shadows that are cast by the inner disc as in \cite{Montesinos+2016}. The star has an effective temperature of $T_{\rm eff} = 5700$ K. The radiative transfer model is set up in spherical coordinates $(R, \theta, \phi)$ with $N_R = 128$, $N_\theta = 90$ and $N_\phi = 64$ grid points in the radial, azimuthal and polar directions respectively. We use a number of photon packages equal to 1 million for our calculations. We run {\sc radmc-3d} simulations at a wavelengths of 1.6 $\mu$m and 1.3 mm in order to model the emission of micron- and mm-sized particles respectively. In Fig.~\ref{fig:obs-radmc}, we report the images obtained for our disc model, after 10\,000 years of evolution.

	\subsubsection{Synthetic ALMA images with {\sc casa}}
	\label{sec:casa}
		
For direct comparison with recent ALMA observations of transition discs, we produce synthetic ALMA observations with the {\sc casa} package \citep{McMullin+2007}. To do so, we scale the radiative transfer calculations of Section~\ref{sec:radiative-transfer} to a distance of 140 pc from the Earth. The bandwidth is 8 GHz and the integration time is 5 hours. We use the ``alma.cycle15.cfg'' array configuration. With this setup we obtain a synthetic beam of 0.08 arcsec$^2$. In Section~\ref{sec:synthobs} we report the synthetic ALMA images obtained. Finally, in Section~\ref{sec:detectability} we discuss the detectability of the dusty spirals in our disc.

%--------------------------------------------------------------------

\section{Results}
\label{sec:results}

In this work we consider two simulations: Sha-$r_{100}$ and NoSha-$r_{100}$, which correspond to the disc model described in Section~\ref{sec:setup} with and without shadows. In both cases, self-gravity (SG) is included. This allows us to disentangle shadowing effects from SG.

	\subsection{Spiral formation and evolution}
	\label{sec:spiralformation}
	
\begin{figure*}
\centering
\includegraphics[width=\textwidth]{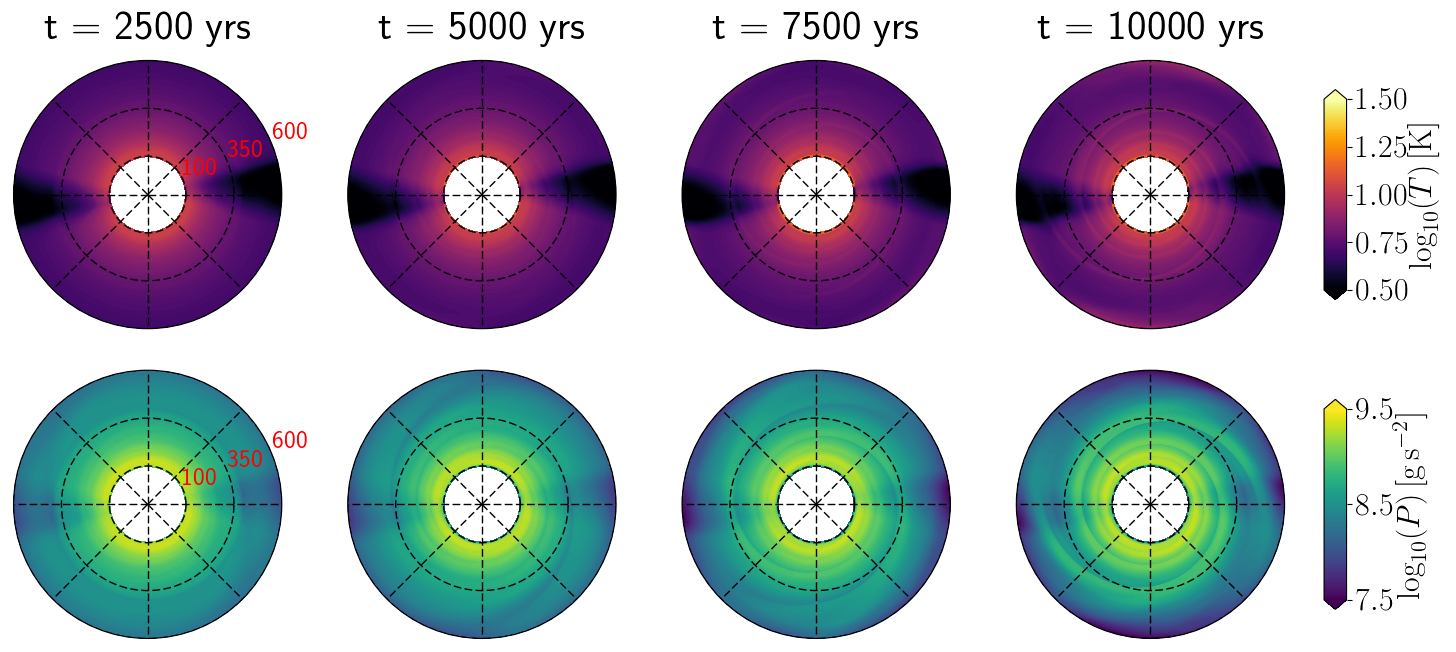}
\caption{Snapshots of the mid-plane gas temperature (top) and the vertically integrated pressure (bottom) at four different snapshots of the Sha-$r_{100}$ simulation: (from left to right) 2\,500, 5\,000, 7\,500 and 10\,000 years.}
\label{fig:disc-evo}
\end{figure*}	

\begin{figure*}
\centering
\includegraphics[width=\textwidth]{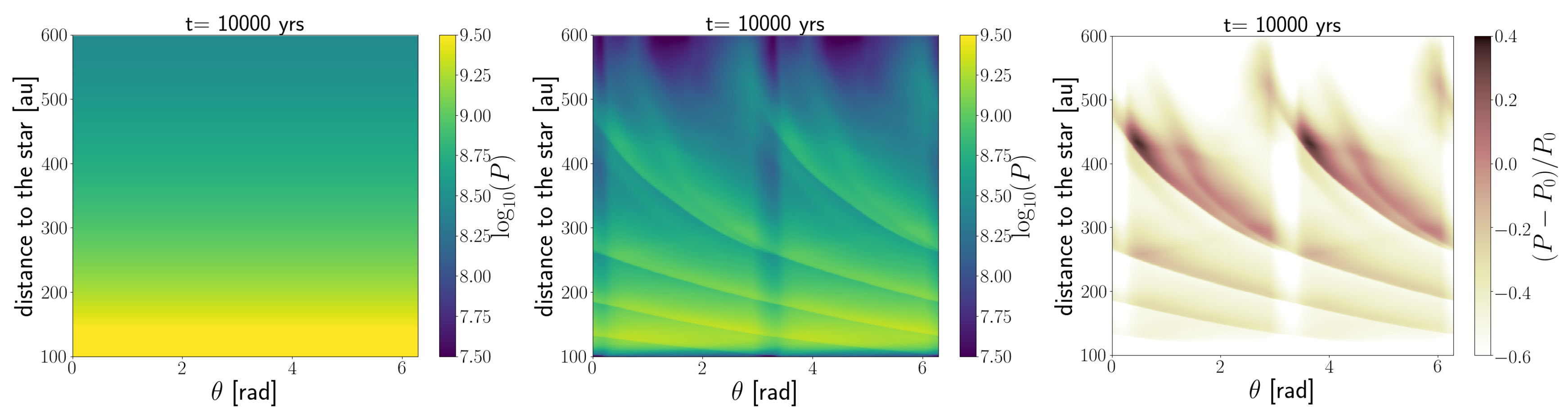}
\caption{$(r,\theta)$ maps of the vertically integrated pressure in NoSha-$r_{100}$ ($P_0$, left), in Sha-$r_{100}$ ($P$, middle) and the normalized perturbation amplitude $(P-P_0)/P_0$ (right). All the snapshots are taken after 10\,000 years of evolution.}
\label{fig:p0pdp}
\end{figure*}	

In Fig.~\ref{fig:disc-evo} we show snapshots at four different times of the Sha-$r_{100}$ simulation. Initially, the surface density of the disc is azimuthally symmetric by construction. In the temperature maps of the disc, the shadowed regions of the disc can be easily seen at azimuths $0\degree$ and $180\degree$. The disc is colder in the shadowed regions because the stellar light is blocked by the inclined inner disc. The cold regions of the disc do not exactly overlap with the regions under the shadows. This effect is due to the anti-clockwise rotation of the disc and the time it takes the material to reach thermal equilibrium. For instance, in the inner regions, the gas crosses the shadows at high angular velocities and hence remains hotter compared to the gas of the outer regions at the same azimuth. This also explains why the shadows' interfaces are smoother than the ones defined by Eq.~\ref{eq:sfunc}. This is also seen in the pressure map, where the shadowed regions have lower values compared to illuminated regions. After 4\,000 years, faint grand-design spirals appear in the pressure and temperature distributions, extending from 200 au up to roughly 400 au. These asymmetries increase in contrast with time as can be seen for further snapshots (7\,500 and 10\,000 yrs), where the spirals extend from 150 up to 500 au.

In clear contrast, the NoSha-$r_{100}$ simulation does not show grand-design spirals in the gas distribution for comparable time-scales. In Figure~\ref{fig:p0pdp}, we show the $(r,\theta)$ pressure maps for NoSha-$r_{100}$ and Sha-$r_{100}$ (left and middle panels respectively). To easily identify the differences, we plot in the rightmost panel the normalised perturbation amplitude $(P-P_0)/P_0$. In the darker regions the pressure is $20\%$ to $40\%$ higher compared to the unperturbed case. As will be shown in Sec.~\ref{sec:dustdynamics}, these local pressure maxima are able to trap dust. On the contrary, brighter regions are characterised by lower pressures. We see for instance that the shadows create two bright lanes at $\theta=0\degree$ and $\theta=180\degree$. Hence, the shadows trigger spirals in the disc, which then constitute regions of high pressure compared to the rest of the disc. Although we take into account self-gravity effects, the disc model considered in this study is gravitationally stable (cf. Appendix~\ref{sec:sanity-checks}). It is worth noting however that, for self-gravitating discs, spirals appear faster in the presence of shadows compared to the case without shadows \citep{Montesinos+2016}. In other words, the shadows render the disc more prone to fragment. To make sure that the shadows are indeed causing the spirals in the disc, we performed a control run where we artificially turned off the self-gravity. In this case, the shadow-triggered spirals still form in the disc proving that self-gravity is a sub-dominant effect.

In our models, the spiral pattern locally co-rotates with the disc, as in the case of self-gravity induced spirals. This contrasts with planetary-induced spirals, which follow the Keplerian frequency of the planet in solid rotation. Due to the quasi-Keplerian motion of our spirals, the dust and the pressure over-density travel at almost the same angular speed, making it easier for the dust to concentrate in the local pressure maxima (cf. Section~\ref{sec:dustdynamics}).

As a side remark, we note that the number of spirals arms in the outer disc is correlated to the number of shadows cast. In fact, by artificially setting one or more than two shadows in the simulations, we observe an equal number of spiral arms and shadows after a few thousands of years. The structure and the evolution of each individual spiral arm remain however the same. These spiral-shaped over-densities in the gaseous distribution constitute a ``sweet spot'' for dust trapping. The efficiency of this mechanism is detailed below.
	
	\subsection{Dust particles dynamics}
	\label{sec:dustdynamics}

\begin{figure*}
\centering
{\includegraphics[width=1\textwidth]{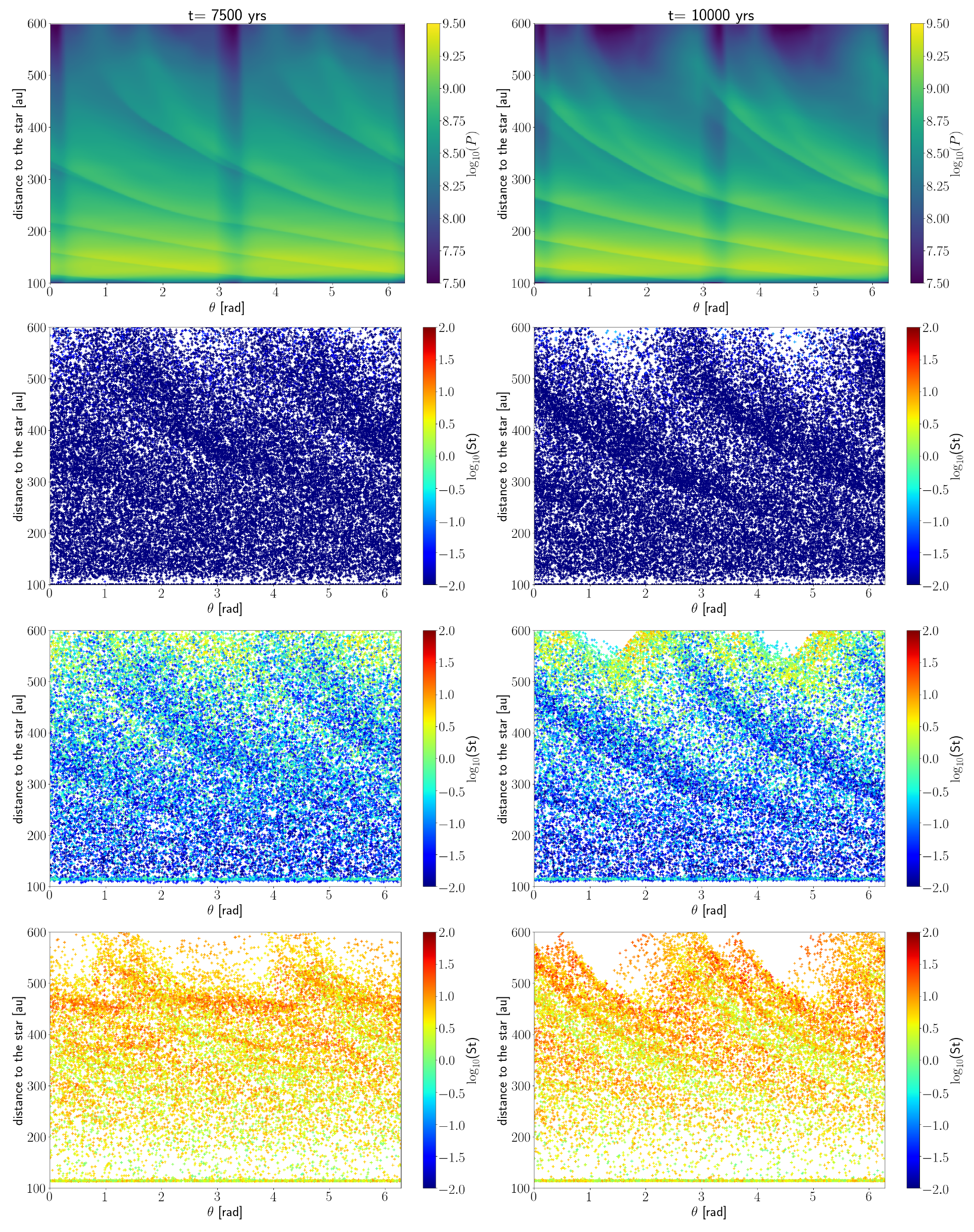}}
\caption{Gas and dust distributions in the ($r$,$\theta$)-plane after $7\,500$ (left) years and $10\,000$ (right) of evolution. The top row shows the vertically integrated pressure map of the disc. The rows below show the dust distribution coloured with the logarithm of the Stokes number (from top to bottom): 100 $\mu$m, 1 mm, 1cm. Spirals in the gas and in the dust have the same morphology.}
\label{fig:rtheta}
\end{figure*}

Figure~\ref{fig:rtheta} shows the distribution of dust, after 7\,500 and 10\,000 years, obtained with the dust evolution code described in Sec.~\ref{sec:dustcode}. We also show the pressure map to compare the location of the dust spirals with the local pressure maxima. We extract three particular sizes from the dust simulation as three different dust species: $a\approx100$~$\mu$m, $a\approx1$~mm, $a\approx1$~cm. Given the different sizes considered, each species has its own dynamical behaviour (cf. Eq.~\ref{eq:St}): the larger the particles, the larger their Stokes number. In the outer parts of the disc, the Stokes numbers are initially close or below $1$. In this case, the larger the particles, the faster the drift towards the pressure maxima. These are located inside the spirals and in the inner region of the disc. This can be seen in Figure~\ref{fig:rtheta}, where in the outer parts of the disc the centimetric particles drift efficiently towards the spirals. The centimetric particles below $r\approx300$ au, being far from the spirals, drift inwards and concentrate at the disc inner edge. Since the pressure (or equivalently the density) inside the spirals is higher than the local mean pressure, the Stokes number is expected to decrease. This effect can be seen in Figure~\ref{fig:rtheta} for centimetric and millimetric particles. In fact, particles with Stokes numbers close to $1$ are efficiently trapped inside the pressure maxima. Consequently, the regions in between the spiral arms get depleted in dust with time. Hence, on longer time-scales, the dust distribution is expected to efficiently concentrate inside the shadow-induced spirals.
    
\begin{figure*}
\centering
\includegraphics[width=0.8\textwidth]{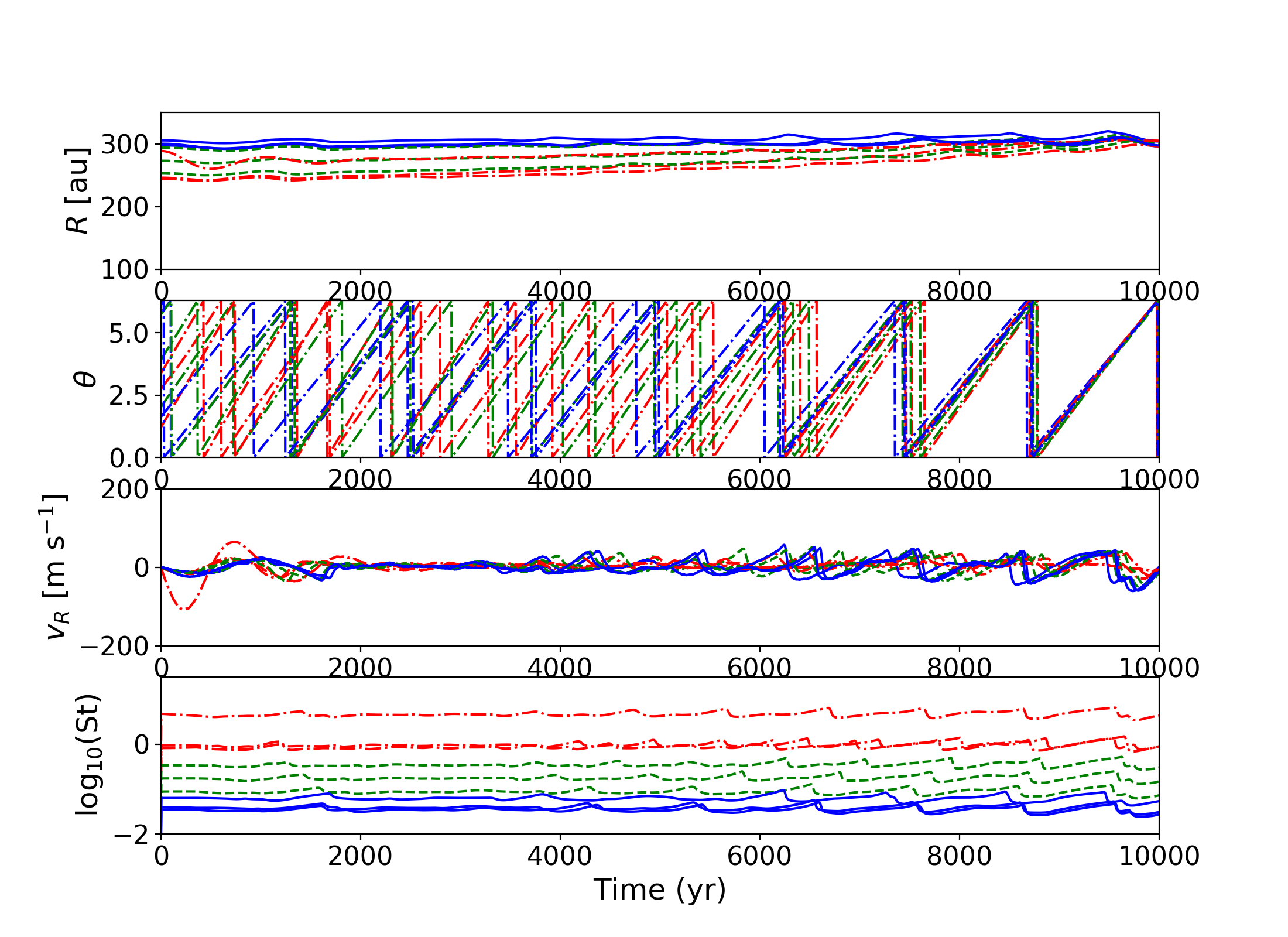}
\caption{From top to bottom: radial and azimuthal positions, radial velocity, and Stokes numbers as a function of time for 3 groups of particles, which end at $r=300 \pm 5$ au and $\theta=0 \pm 5$. We plot 100 $\mu$m particles in blue, 1 mm particles in green and 1 cm particles in red. Each group of particles contains 3 particles of equal size.}
\label{fig:trajr300}
\end{figure*}

\begin{figure*}
\centering
\includegraphics[width=0.8\textwidth]{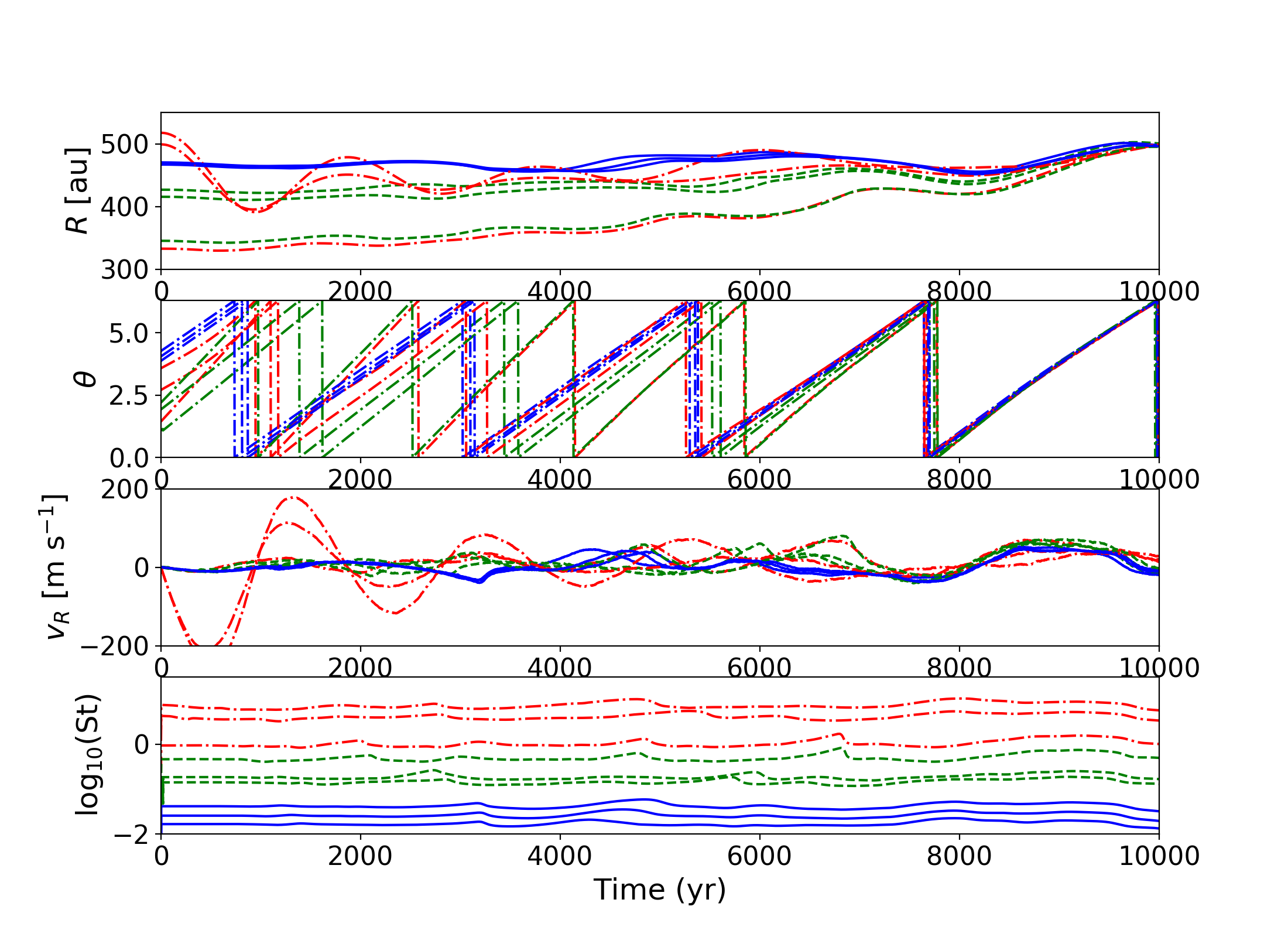}
\caption{Same as Fig.~\ref{fig:trajr300}, but for $r=500 \pm 5$ au.}
\label{fig:trajr500}
\end{figure*}

In order to study the characteristic behaviour of each species, we select individual but representative particles at a specific radial and azimuthal position of the disc at the end of the simulation. Then, we trace back their evolution in the disc to find their initial positions. We chose two sets of particles: one at $r=300 \pm 5$ au and another at $r=500 \pm 5$ au, both at $\theta=0 \pm 5\degree$. It is worth noting that taking $\theta = 180 \pm 5$ we find similar results. Figures~\ref{fig:trajr300} and \ref{fig:trajr500} show the distance to the star, the azimuth, the radial velocity and the Stokes numbers as a function of time for particles of different sizes belonging to each set. We show three representative particles for each size bin: 100 $\mu$m in red, 1 mm in green and 1 cm in blue. By selection, the particles plotted have the same final radial and azimuthal positions. However, these were different at $t=0$ yrs. In fact, in Fig.~\ref{fig:trajr300}, we can see that the particles with small Stokes numbers (blue) remain at roughly the same radial position during the whole simulation, while those with St$\approx$1 (red) experience an outward movement. Particles with St$\approx$0.1 show the same behaviour with a less rapid outward drift. Similar but more pronounced effects are observed for particles at $r=500$ au at the end of the simulation. This is due to the greater pressure gradients present in the disc at larger stellocentric distances (cf. Fig.~\ref{fig:p0pdp}). In addition, we see that some particles with St$\approx$1 beyond $500$ au at $t=0$ initially drift inwards and then concentrate at 500 au. This phenomenon is consistent with dust drifting towards\footnote{both from larger and smaller radii.} the pressure maximum located at the spiral position. This is in contrast with what is observed for an unperturbed disc (i.e. without shadows) where all the dust drifts inwards (cf. Appendix~\ref{sec:benchmark}). Moreover, the radial velocity is in agreement with the dynamical behaviour described. In fact, we observe that this velocity (typically negative) becomes slightly positive because of the presence of the pressure maxima. Also, it shows small oscillations due to the shadows' crossing. Hence, the gaseous spirals act as efficient dust traps gathering material of varied sizes, starting at different radial positions. 

	\subsection{Synthetic observations}
	\label{sec:synthobs}

\begin{figure*}
\centering
{\includegraphics[width=0.48\textwidth]{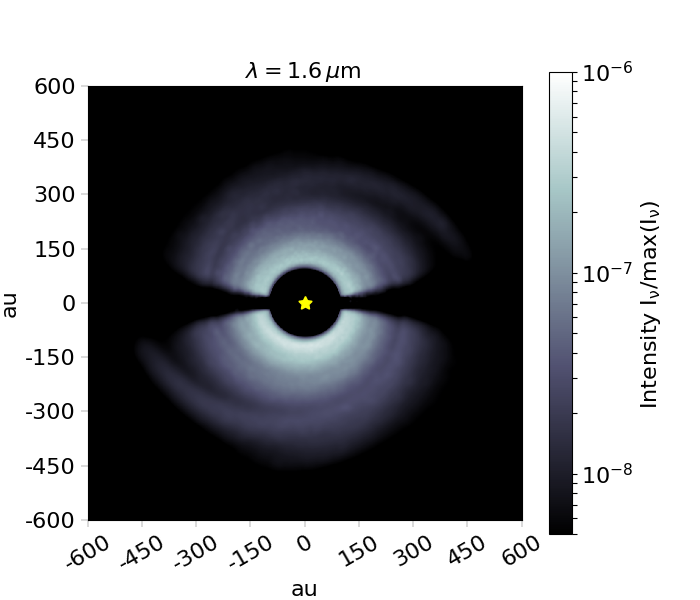}}
{\includegraphics[width=0.48\textwidth]{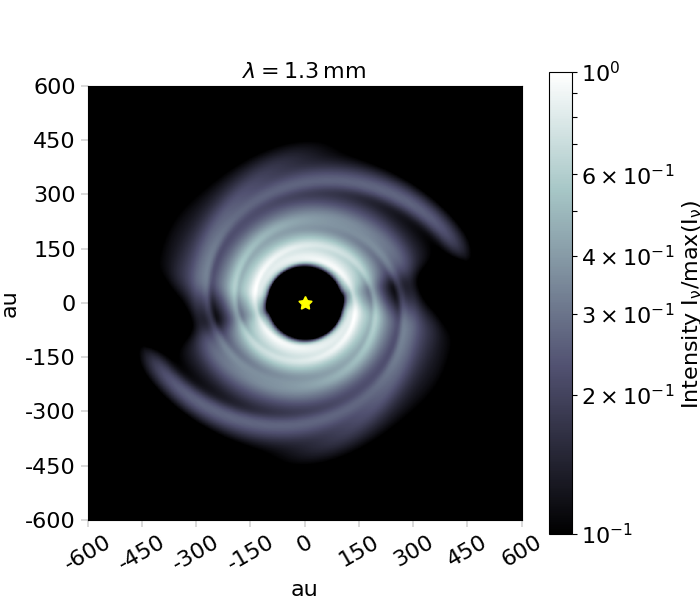}}
\caption{Dust thermal emission at 1.6 $\mu$m (left) and 1.3 mm (right) after 10\,000 years obtained with {\sc radmc-3d}. The source is placed at $d=140$ pc and the disc is seen almost face-on ($i=13\degree$).}
\label{fig:obs-radmc}
\end{figure*}

Figure~\ref{fig:obs-radmc} shows the radiative transfer calculation for Sha-$r_{100}$ after 10\,000 years of evolution at $\lambda_1 = 1.6 \,\mu$m and $\lambda_2 = 1.3$ mm obtained following the steps in Section~\ref{sec:radiative-transfer}. As mentioned in Section~\ref{sec:intro}, the former traces the small dust and the latter the millimetric grains. The value of $\lambda_2$ was chosen in order to have the same wavelength as the observations reported by \cite{Perez+2016}. We assume the system is 140 pc away from the Earth and has an inclination equal to $13\degree$, i.e. almost face-on. Both images clearly show the two diametrically-opposed shadows, which are expected by construction. In addition, faint spiral patterns are seen at $\lambda_1 = 1.6 \,\mu$m, while these appear sharper and more radially concentrated at $\lambda_2 = 1.3$ mm. The combined effect of radial-drift and dust trapping inside the spirals explains the difference between both images. We should stress however that the thermal emission from the disc inner rim at roughly $100$ au is at least one order of magnitude brighter than the one emitted by the spirals arms. This has dramatic consequences for the detectability of such thermal emission.

In Figure~\ref{fig:casa} we show the post-processed image with {\sc casa}, which emulates an ALMA observation (cf. Section~\ref{sec:casa}). In particular, the presence of two dips at $\theta = 0\degree$ and $\theta = 180\degree$ shows that it is in principle possible to detect the shadows at millimetric wavelengths. This is in good agreement with the ALMA synthetic images obtained by \cite{Facchini+2018}, where an inclined inner disc cast shadows onto the outer disc. Unfortunately, in our case, due to the low signal to noise in the spirals it is not possible to detect any significant thermal emission from the dusty spiral arms. Even considering longer integration times and/or other antenna configurations, the dusty spirals remain below the detection threshold.
    	
%-----------------------------------------------------------------------------------------

\section{Discussion}
\label{sec:discussion}

	\subsection{Grand-design dusty spirals}
	\label{sec:dustyspirals}
	
We considered a new mechanism able to produce grand-design dusty spirals in transition discs. Remarkably, there is no need to include a massive planetary or stellar companion as in \cite{Dong+2015}. We only require a couple of diametrically opposed shadows cast at the inner rim of a transition disc as in \cite{Montesinos+2016} and \cite{Facchini+2018}. The resulting gaseous grand-design spirals are able to trap dust of different sizes in the disc. Our calculations show that the dust particles with St$\sim 1$ are the ones that gather most efficiently in the spirals. Interestingly, these particles move towards the pressure maxima both from inwards and outwards. This effect is not seen in the absence of shadows and the resulting spirals. Despite the similar morphology of our dusty spirals (cf. Fig.~\ref{fig:obs-radmc}) with the ones observed by \cite{Perez+2016}, our synthetic ALMA image shows that these shadow-triggered spirals cannot be detected at $1.3$ mm. Hence, shadows are a promising mechanism to produce dusty spirals in disc but it is not possible to reproduce the observations of Elias 2-27. Moreover, \cite{Perez+2016} do not report any shadow detection. However, this mechanism should be considered for other transition discs exhibiting large cavities and shadows (e.g. HD~142527, HD~135344B and HD~100453).
	
It is worth noting that we studied the evolution of shadow-triggered spirals for only 10\,000 years, a time-scale too short to consider grain growth and fragmentation inside the spirals. Although our dust evolution code has a grain growth module, the typical grain growth time-scales in the region of interest exceed the simulation time by at least one order of magnitude. That is the reason why we neglected this phenomenon, even though we expect particles to collide and grow inside the spirals. Unfortunately, simulating the disc for a prolonged time is problematic due to the large number of snapshots we need to store in the memory.

	\subsection{Dust trapping and mixing}
	\label{sec:dustmixing}
	
In Section~\ref{sec:dustdynamics}, we analysed the dust trapping by the shadow-triggered spirals in the gas. This phenomenon depends on the grain size and influences the dust dynamics in the disc dramatically. Figure~\ref{fig:rtheta} clearly shows that dust particles concentrate in the pressure maxima of the gas, which results into a dusty disc with spirals. The fact that particles are trapped in these over-densities, prevents these solid bodies to be accreted by the central star. In fact, in the absence of spirals, only particles with St $\sim$ 1 experience an extremely fast radial-drift \citep{W77}. Eventually, these particles are lost onto the star, which renders planetesimal formation rather problematic in the disc. Because inside the spirals dust particles can gather and grow efficiently, these constitute remarkable ``sweet spots'' for grain growth in the disc. This gathering effect could lead to streaming instabilities at this particular location \citep{Youdin&Goodman2005, Johansen+2007, Jacquet+2011}, but this effect has not been taken into account in our simulations. Provided that strong dust clumping happens, the radial-drift is slowed down in these over-dense regions. Once the self-gravity of the solids becomes important, this mechanism can lead to the formation of planetesimals \citep{Johansen+2007, Johansen+2012}.

In addition to this effect, we expect an efficient mixing of dust species inside the spirals. In Figures~\ref{fig:trajr300} and \ref{fig:trajr500} we observe a mixing of particles inside the spiral at $\sim$300 au and at $\sim$500, which initially started at different radial and azimuthal distances. In fact, particles with different initial Stokes number drift at different radial velocities until they are trapped in the over-density. This happens both in the inward and outward direction. Once this happens, there is a broad dust size distribution in the spiral. The fact that the trapped grains come from regions of the disc at different temperatures may result in different solid compositions (ices, volatile organics, refractory organics, silicates, iron, etc.). The mixing among different species is controlled by the numerator of Eq.~\ref{eq:St}, $\rho_{\rm}$ and $a$, as discussed in \cite{Pignatale+2017}. In fact, a dense but small particle ($\rho_{\rm d}=10$ ${\rm g \, cm^{-3}}$, $a=0.1$ mm) behaves in the same way as a lighter but larger particle ($\rho_{\rm d}=1$ ${\rm g \, cm^{-3}}$, $a=1$ mm). This has dramatic effects on the resulting composition of solids in the disc, and also affects their further growth \citep[and references therein]{Pignatale+2017}.
	
However, caution is required when interpreting our results given that we are considering an initial size distribution which does not depend on the radius. If we consider a disc with a realistic size segregation, then the emission from the outer regions at $\lambda=1.3$ mm would be fainter and the one from the inner regions stronger. In fact, due to radial-drift, we expect to have a more compact radial distribution of millimetric grains compared to micron-sized ones.
	
\subsection{Detectability of shadow-triggered dusty spirals}
	\label{sec:detectability}

\begin{figure}
\centering
{\includegraphics[width=0.5\textwidth]{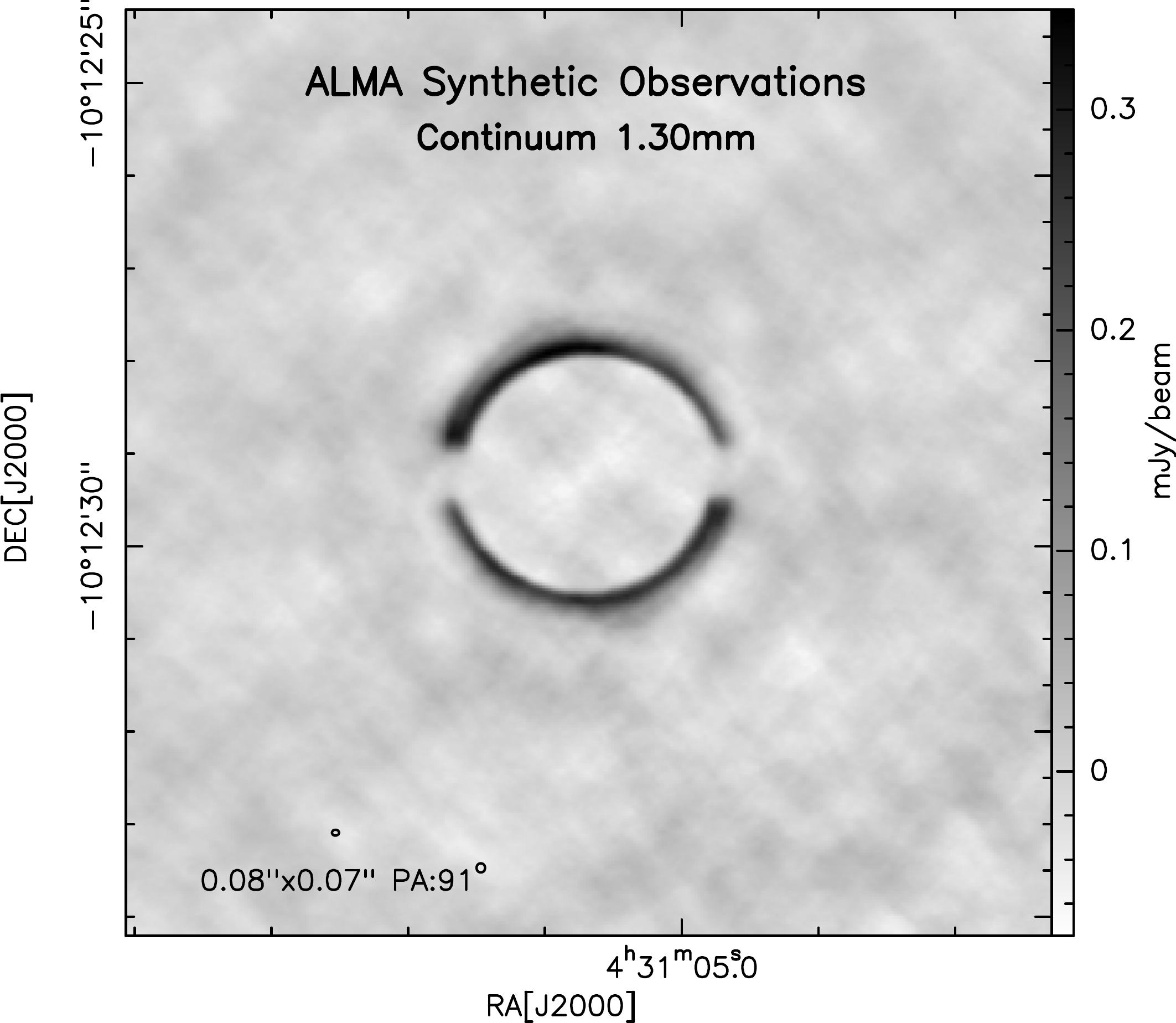}}
\caption{Synthetic observations with ALMA of Sha-$r_{100}$ after 10\,000 years. The beam is indicated in the bottom-left hand side.}
\label{fig:casa}
\end{figure}

In Figure~\ref{fig:casa} we report a synthetic ALMA observation corresponding to Sha-$r_{100}$. Although the shadows are clearly detected as dips in the emission at $1.3\,$mm, the thermal emission from the spirals is comparable to the noise level. Hence, even for longer integration times, their emission will remain below the detection limit. Unrealistically higher disc masses or dust-to-gas ratios could increase the thermal emission rendering such detections feasible. If the canonical value of $1\%$ of the latter is in fact too conservative, then shadow-triggered spirals could potentially be observable. The size of the cavity also plays an important role. As a matter of fact, discs with smaller cavities have higher fluxes because of the presence of hotter material closer to the star. This  decreases the flux ratio between the spiral arms and the inner disc, which makes the detection even more challenging. Consequently, transition discs exhibiting shadows are ideal targets to look for spiral patterns in the thermal emission. It is worth noting nevertheless that considering a lower disc mass (e.g. $1\%\,M_{\odot}$), would produce even fainter spirals. Therefore, the disc model presented in this work should be seen as an optimistic scenario for this kind of spiral formation and detection.

Interestingly, some recent techniques allow to extract structure from noisy data well below the 5$\sigma$-threshold. For instance, in the context of galaxy observation, \cite{Akhlaghi+2015} describe a method able to extract spiral arms at very low signal-to-noise ratio (SNR). Contrary to other existing techniques, signal-related parameters (such as the image point spread function or known object models) are irrelevant for this method. Thus, assuming that this kind of dusty spirals form in Nature, some previous (sub-)millimetric detections may have missed them because of standard data reduction procedures. To our knowledge, this kind of observational signatures has not been observed with ALMA yet. However, based on our results, we are able to predict that, in presence of shadows, the thermal emission of the dust should exhibit faint grand-design spirals and significant dips (cf. Figure~\ref{fig:obs-radmc}).

%--------------------------------------------------------------------

\section{Conclusions}
\label{sec:conclusions}

In this work, we have developed a novel numerical tool, compatible with {\sc fargo-adsg}, to explore dust dynamics for hydrodynamical simulations in a consistent manner. Although we do not include the back-reaction of dust on gas, this does not affect our results given the dust-to-gas ratios considered here. The methodology detailed in Section~\ref{sec:method} allows to compute the evolution of the dust distribution and its resulting thermal emission. In this regard, this tool successfully bridges the gap between theory and observations. In fact, it is particularly well-suited to model some of the transitional discs exhibiting puzzling dust distributions (cf. Table~\ref{tab:obs}). 
	
With the detection of dusty spirals in Elias 2-27 \citep{Perez+2016} and the mechanism described in \cite{Montesinos+2016} in mind, we explored the trapping of dust in discs with gaseous spirals due to illumination effects. In our case, the shadows are provoked by an inclined inner disc, which obstructs the stellar light at large radii. The main results of this work can be summarised as follows:
\begin{itemize}
\item The gaseous spiral waves induced by shadows are able to accumulate and trap dust particles. This mechanism deeply affects the fate and the evolution of the solid material in the disc. In fact, we showed that for mm to cm-sized particles it is possible to circumvent the radial-drift barrier.
\item In our simulations, the most efficient trapping occurs for solids with sizes ranging from 1 mm to 1 cm. However, this size range strongly depends on the disc model considered \citep{Laibe+2012}.
\item The dust trapping that occurs inside the gaseous spirals results in an efficient mixing of dust particles of different sizes and densities. This has deep implications for planetesimal and comet formation at large stellocentric distances, since spirals gather material from different regions of the disc.
\item The obtained infrared (1.6 $\mu$m) and thermal (1.3 mm) emissions exhibit both shadows and spiral patterns in the disc. However, the brightness of the dusty spirals is too faint to explain the recent ALMA observations of Elias 2-27.
\end{itemize}

The methodology presented here can be applied to explore dust dynamics in transition discs with vortices, multiple planets and moving shadows. By doing so, it is possible to produce robust observational signatures of different physical processes at act. This will hopefully help observers to interpret their current and future dust detections.

\begin{acknowledgements}
We thank the two anonymous referees that have reviewed this manuscript for very constructive comments, which allowed us to enhance the quality of this work. NC acknowledges financial support provided by FONDECYT grant 3170680. NC, MM, and JC acknowledge financial support from Millenium Nucleus grant RC130007 (Chilean Ministry of Economy). MM acknowledges support from the Millennium Science Initiative (Chilean Ministry of Economy) and the CHINA-CONICYT fund, 4th call. SMS gratefully acknowledges support through the PUC-HD Graduate Student Exchange Fellowship, which is part of the academic exchange program between the Institute of Astrophysics of the Pontificia Universidad Cat\'olica (IA-PUC) and the Center for Astrophysics at the University of Heidelberg (ZAH), financed by the German Academic Exchange Service (DAAD). This work was partly carried out while JC was on sabbatical leave at MPE.  JC and NC acknowledge the kind hospitality of MPE, and funding from the Max Planck Society through a ``Partner Group'' grant.  JC acknowledges support from CONICYT-Chile through FONDECYT (1141175) and Basal (PFB0609) grants, and from the ICM (Iniciativa Cient\'ifica Milenio) via the N\'ucleo Milenio de Formaci\'on Planetaria grant. The Geryon/Geryon2 cluster housed at the Centro de Astro-Ingenieria UC was used for the calculations performed in this paper. The BASAL PFB-06 CATA, Anillo ACT-86, FONDEQUIP AIC-57, and QUIMAL 130008 provided funding for several improvements to the Geryon/Geryon2 cluster.
\end{acknowledgements}

\bibliographystyle{aa}
\bibliography{astro}

\begin{appendix}

\section{Irradiation versus accretion heating}
\label{sec:heating}

Here we aim to show that we can safely neglect the effect of the viscous heating in the disc. On the one hand, we consider the viscosity of the gaseous disc, which leads to the production of heat. The heating per gram of gas is proportional to the the viscosity coefficient $\nu$ and the square of the shear. In an accretion disc, the total heating per unit area, $Q_+^{\rm accr}$, reads:
\begin{equation}
Q_+^{\rm accr} = \Sigma \nu \left( r \, \frac{{\rm d}\Omega_{\rm K}}{{\rm d}r} \right)^2 = \frac{9}{4} \Sigma \nu \Omega_{\rm K}^2 \,\,\, .
\end{equation}
Assuming a steady state disc, the accretion rate can be written as follows: $\dot M = 3 \pi \Sigma \nu$. Thus, we obtain:
\begin{equation}\label{eq:Qvisc}
Q_+^{\rm accr} = \frac{3}{4 \pi} \dot M \Omega_{\rm K}^2 = \frac{3 G}{4 \pi} \frac{\dot M M_*}{r^3} \,\,\, .
\end{equation}
On the other hand, the stellar heating directly depends on the amount of radiation received by the disc. The flux of stellar radiation at distance $r$ from the star is given by $F_*(r) = \frac{L_*}{4 \pi r^2}$. The irradiating flux is the projection of this flux onto the surface of the disc and includes a factor $\sin(\phi)$, which can be approximated by $\phi$ for small grazing angles. Typically, for a flared disc, $\phi \approx 0.05$. The disc receives radiation from the upper and the lower part, therefore there is a factor 2 as well. Finally, we get that the irradiation heating per unit area, $Q_+^{\rm irr},$ is given by:
\begin{equation}\label{eq:Qirr}
	Q_+^{\rm irr} = \phi \frac{L_*}{2 \pi r^2} \,\,\, .
\end{equation}
Hence, the criterion for which the irradiation heating dominates over the accretion one reads as follows:
\begin{equation}\label{eq:heat-criterion}
	r \geq \frac{3 G \dot M M_*}{2 \phi L_*} \,\,\, .
\end{equation}
Plugging a reasonable value for the stellar accretion ($\dot M = 10^{-7} M_*$), we obtain that $r \geq 11$ au. This condition is fulfilled by our disc, so the viscous heating can be neglected.

%
% ------------------------------------------------------------------------------------------------------------------------------
%

\section{Disc temperature and Toomre parameter}
\label{sec:sanity-checks}

\begin{figure}
\centering
\includegraphics[width=0.5\textwidth]{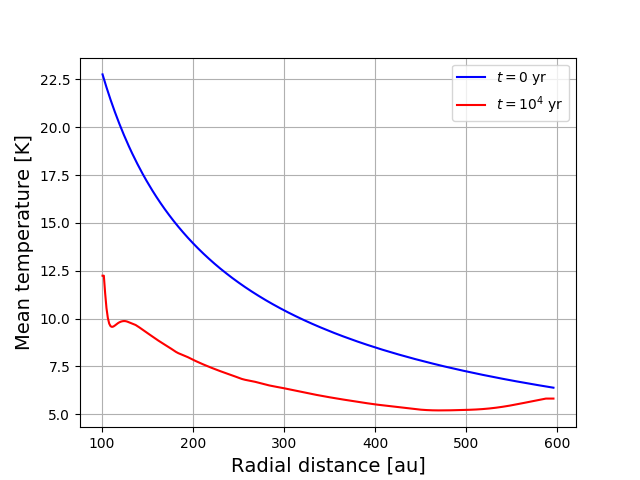}
\caption{Initial and final radial temperature profiles for the Sha-$r_{100}$ simulation. The range of temperature is consistent with Eq.~\ref{eq:Temp}.}
\label{fig:temperature}
\end{figure}

\begin{figure}
\centering
\includegraphics[width=0.5\textwidth]{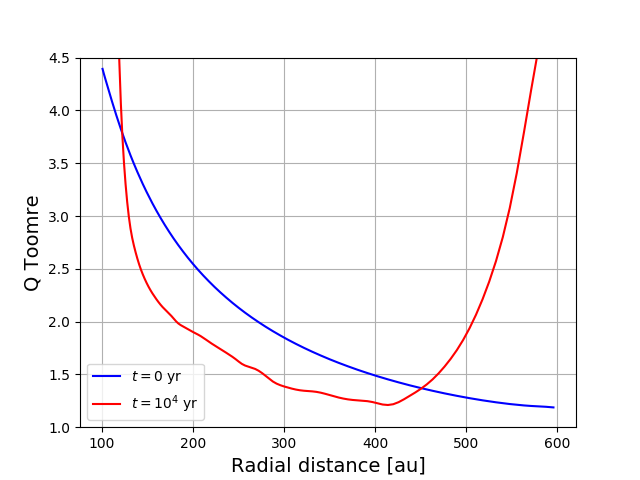}
\caption{Initial and final radial Toomre parameter profiles for the Sha-$r_{100}$ simulation. Values above $1$ correspond to gravitationally stable discs. This is the case for the disc model considered in this work.}
\label{fig:toomre}
\end{figure}

In Figs.~\ref{fig:temperature} and \ref{fig:toomre} we show the initial and final values of the disc temperatures and the Toomre parameter, respectively. Because the disc radiates energy during its evolution, the final temperatures are lower than the initial ones and are roughly comprised between 12 and 5 K. This is consistent with the expected temperatures between 100 and 600 au for an irradiated disc (cf. Eq.~\ref{eq:Temp}). The Toomre parameter is defined as $Q=c_{\rm s} \kappa_{\rm f} / \pi G \Sigma$, where $\kappa_{\rm f}$ is the epicyclic frequency. This dimensionless quantity compares the strength of thermal pressure and shear (numerator) to self-gravity (denominator). If $Q>1$ ($Q<1$), then the system is (un)stable to its own gravity. In our simulations, the value of the Toomre parameter throughout the entire evolution remains everywhere above $1$. Regions with $Q$ values slightly above $1$, which are found in our model, can be considered as marginally stable \citep{KratterLodato2016}. However, even in that regime, the perturbation introduced by the shadows dominates over self-gravity effects, leading to spiral formation \citep[see][for a more thorough discussion]{Montesinos+2016}.

\section{Numerical integration of the equations of motion}
\label{sec:integration}

As mentioned in Section~\ref{sec:dustcode}, the expression of the Stokes number used in the code is different than Eq.~\ref{eq:St}. In the dust code, we use the radial and the azimuthal gravitational accelerations given by {\sc fargo-adsg} in order to compute the Stokes number. For a dust particle of index $i$, we have:
\begin{equation}
\label{eq:St-code}
	{\rm St}_i = \sqrt{\frac{\pi}{2}} \frac{(3 {\rm Kn} + 1)^2}{9{\rm Kn}^2 + g_{\rm drag} + 3 {\rm Kn} \, k_{\rm drag}} \frac{a}{r_i} \frac{\zeta_{\rm dust}}{\rho_{\rm gas}} \frac{v_{\rm K}}{c_{\rm s}} \,\, ,
\end{equation}
where ${\rm Kn} = \lambda_{\rm mfp} / a$ is the Knudsen number, $\lambda_{\rm mfp}$ the mean free path of the gas and $a$ the radius of the dust particle. $\zeta = 3.3 \, {\rm g \, cm^{-3}}$ is the bulk mass density of the dust particles, and $\rho_{\rm gas}$ the mid-plane mass density of the gas. $g_{\rm drag}$ and $k_{\rm drag}$ are the drag coefficients in the Epstein and in the Stokes drag regimes, respectively. They are given by:
\begin{equation}
	g_{\rm drag} = \sqrt{1+\frac{9 \pi}{128} m^2}
\end{equation} 
and
\begin{equation}
    k_{\rm drag} = 
  	\begin{cases}
      1 + 0.15 \, {\rm Re}^{0.687} & \text{if}\ {\rm Re} \leq 500 \\
      3.96 \times 10^{-6} \, {\rm Re}^{2.4} & \text{if}\ 500 < {\rm Re} \leq1500 \\
      0.11 \, {\rm Re} & \text{if}\ {\rm Re} > 1500.
    \end{cases}
  \end{equation}
${\rm Re}$ is the Reynolds number, which is defined as:
\begin{equation}
	{\rm Re} = 3 \sqrt{\frac{\pi}{8}} \frac{m}{\rm Kn} \,\, , 
\end{equation}
where $m =  | \Delta {\mathbf{v}} | / c_{\rm s}$ is the relative Mach number of the dust particles to the gas. This rather complex method for calculation the Stokes number has the advantage that it automatically interpolates between the different drag regimes and is therefore more stable.

We use a symplectic, first-order leapfrog method to integrate Eqs.~\ref{eq:eomdust1} to \ref{eq:eomdust4}. Equations containing the Stokes number are integrated implicitly. First, Eqs.~\ref{eq:eomdust1} and \ref{eq:eomdust2} are integrated using only half of a time step $\Delta t$. Omitting the particle index $i$, the integration from the $k^{\rm th}$ to the $(k+1)^{\rm th}$ time-step is performed as follows:
\begin{eqnarray}
	r_{k+1/2} &=& r_k + \frac{\Delta t}{2} v_{r,k} \\
	\theta_{k+1/2} &=& \theta_k + \frac{\Delta t}{2} \frac{L_k}{r_k^2} \,\, .
\end{eqnarray}
Then, Eqs.~\ref{eq:eomdust3} and \ref{eq:eomdust4} are integrated with a full time-step and using the new values $r_{k+\frac{1}{2}}$ and $\theta_{k+\frac{1}{2}}$:
\begin{eqnarray}
	v_{r, k+1} &=& \frac{v_{r, k} + \Delta t \left( \frac{L_k^2}{r_{k+1/2}^3} - \frac{\partial \Phi}{\partial r} + f_{\rm acc}, r + \frac{v_{{\rm gas},r} \, \Omega_{\rm K}}{\rm St} \right)}{ 1 + \frac{\Delta t \Omega_{\rm K}}{\rm St} }  \\
	L_{k+1} &=& \frac{L_k + \Delta t \left( - \frac{\partial \Phi}{\partial \theta} + r_{k+1/2} \, f_{{\rm acc}, \theta} + \frac{v_{{\rm gas}, \theta} r_{k+1/2} \, \Omega_{\rm K}}{\rm St} \right) }{ 1 + \frac{\Delta t \Omega_{\rm K}}{\rm St} } \,\, .
\end{eqnarray}
where $f_{{\rm acc}, r}$ and $f_{{\rm acc}, \theta}$ are the gravitational accelerations in $r$ and $\theta$ directions due to the gravitational potential of the disc, extracted from {\sc fargo-adsg}.

Finally, Eqs.~\ref{eq:eomdust1} and \ref{eq:eomdust2} are integrated again with half of a time-step using $L_{k+1}$ and $v_{r,k+1}$ in the following way:
\begin{eqnarray}
	R_{k+1} &=& R_{k+1/2} + \frac{\Delta t}{2} v_{r, k+1} \\
	\theta_{k+1} &=& \theta_{k+1/2} + \frac{\Delta t}{2} \frac{L_{k+1/2}}{R_{k+1/2}^2} \,\, .
\end{eqnarray}

To ensure the accuracy of this method, every quantity is linearly interpolated onto the location of the dust particle. To be stable, the particles perform several time-steps between two {\sc fargo-adsg} outputs. Between two outputs, the gas quantities are linearly interpolated in time. To be as precise as possible, two consecutive {\sc fargo-adsg} snapshots should not lie too far apart. As a rule of thumb, ten or more snapshots per orbit at the inner boundary of the system are enough for this numerical scheme.

Furthermore, a random walk term can be added to simulate the turbulent diffusion of the dust. If activated, after every time-step, the particles are displaced by a length $l_{\rm turb}$ in a random direction. The distance of displacement is given by \cite{Youdin+Lithwick-2007} as:
\begin{equation}
	l_{\rm turb} = \Delta t \, \frac{\alpha H^2 \Omega_{\rm K}}{1 + {\rm St}^2} \,\, .
\end{equation}
This option is turned off in all the simulations presented here.

%
% ------------------------------------------------------------------------------------------------------------------------------
%
	
\section{Benchmark of the dust evolution code}
\label{sec:benchmark}

To prove that the equations of motions are integrated accurately, we performed simulations of several test scenarios. In all cases, we kept the gas parameters constant with a surface density of $\Sigma(r) = 1700 \, {\rm g \, cm^{-2}} \times \left( \frac{r}{1 \, {\rm au}} \right)^{-1}$ and a temperature profile of $T(r) = 150 \, {\rm K} \times \left( \frac{r}{1 \, {\rm au}} \right)^{-1/2}$. We consider a disc orbiting a $1 \, M_\odot$ star. For an axisymmetric disc, the radial evolution of a given dust particle of index $i$ is governed by:
\begin{equation}\label{eq:radial-drift}
	\frac{\partial}{\partial t} r_i = \frac{v_{{\rm gas},r}}{1 + {\rm St}^2} - \frac{2 \, u_{\rm P}}{{\rm St} + {\rm St}^{-1}} \,\, ,
\end{equation}
where $u_{\rm P}$ is the maximum particle drift velocity caused by the sub-Keplerian gas \citep{Birnstiel+2016}. In a fixed gas disc, the particle's radial evolution is solely dependent on the particle's Stokes number.

To check whether the code is stable and correct for a broad variety of Stokes numbers, we performed four test scenarios. In three cases we kept the Stokes number of the particles constant at $10^{-100}$, $10^{100}$ and $1$. In the first case, the particles are completely coupled to the gas and are therefore being accreted with the radial gas velocity $v_{{\rm gas},r}$. In the second case, the particles are completely decoupled from the gas, are on Keplerian orbits and do not change their radial position. In the third case with St = 1, the particles experience the maximum radial drift velocity of $\frac{1}{2} v_{{\rm gas},r} + u_{\rm P}$. We also performed a fourth test case where we did not fix the Stokes number at all. Here, the Stokes number will decrease with time as the particles drift to regions with higher surface densities.

\begin{figure*}
\centering
\includegraphics[width=\textwidth]{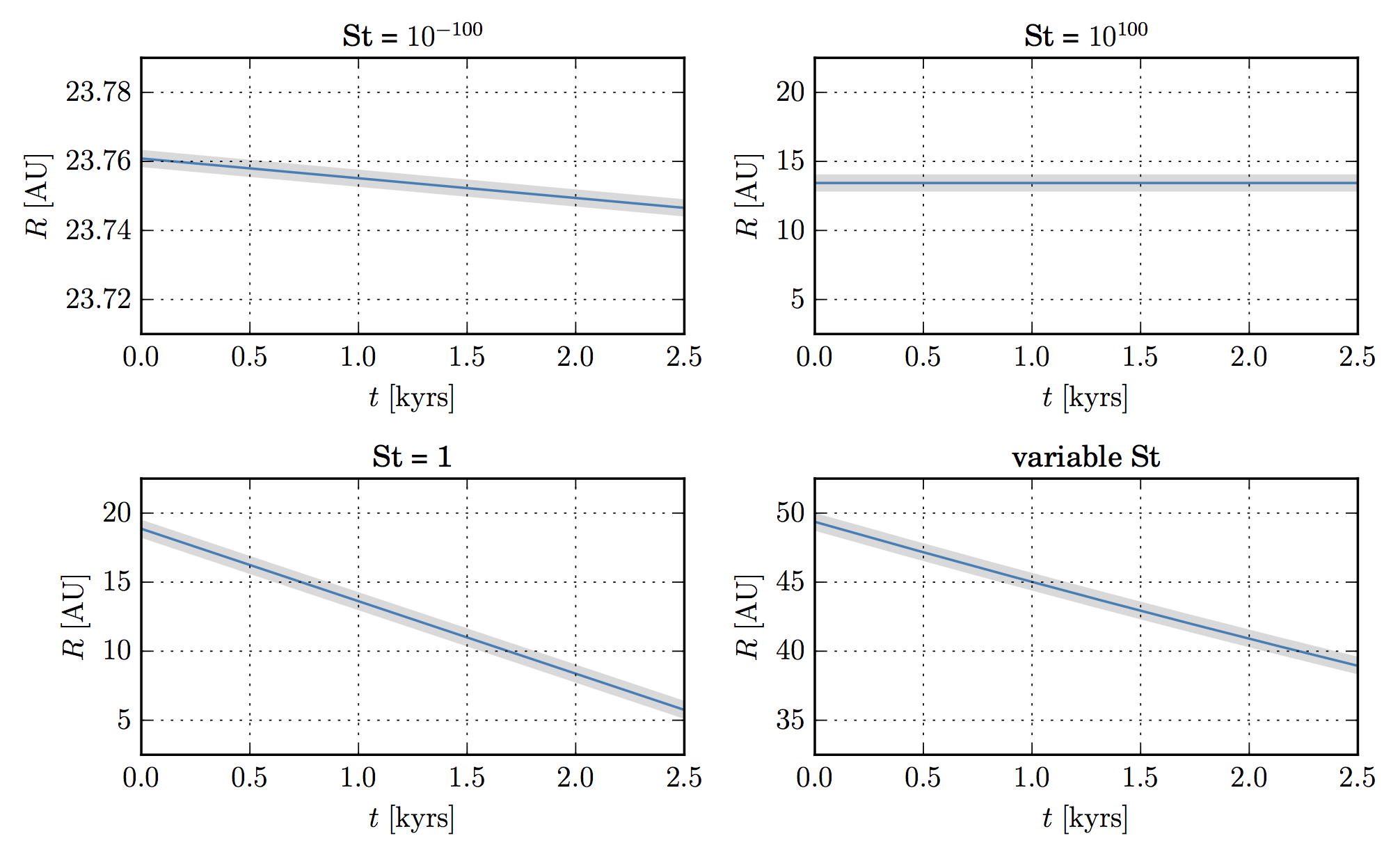}
\caption{Test cases of the particle simulation. Comparison of the radial evolution for test particles in four different scenarios. The results obtained with our dust code are colored in blue, while those obtained through the direct integration of Eq.~\ref{eq:radial-drift} are colored in gray. From top to bottom and from left to right, we plot 4 different cases: ${\rm St} = 10^{-100}$, ${\rm St} = 10^{100}$, ${\rm St} = 1$, and variable Stokes numbers. The gas disc does not evolver in any of these test cases.}
\label{fig:benchmark}
\end{figure*}	

Figure~\ref{fig:benchmark} shows the radial evolution of selected test particles in each of the four scenarios with blue lines. We also directly integrated Eq.~\ref{eq:radial-drift} for all cases and over-plotted the expected particle tracks with broad gray lines. The top left shows a particle with a very small Stokes number. It is perfectly coupled to the gas and is accreted with the radial gas velocity. The particle on the top right has a very high Stokes number and is therefore on an Keplerian orbit. Its radial position does not change at all. The particle on the bottom left has a Stokes number of unity and is accreted with the maximum drift velocity. Due to our choice of the surface density and temperature profiles, this drift velocity is constant with radial distance from the star. The gas velocity can be neglected here. The particle on the bottom right has a variable Stokes number that decreases with decreasing distance from the star due to higher surface densities in the inner disc. Therefore, the particle's radial velocity decreases slightly over time. Both, the results from the particle simulation and the direct integration match perfectly. We therefore conclude that the particle code is stable for all Stokes numbers and returns the correct results.

\end{appendix}

\end{document}